\documentclass[twocolumn]{aastex631}

\usepackage{amsmath}

\begin{document}

\title{LSST Strong Lensing Systems Dark Matter Sensitivity Analysis with Neural Ratio Estimators}

\author[0000-0003-4701-3469]{Andreas Filipp}\email{andreas.filipp@umontreal.ca}
\affiliation{Ciela Institute, Montreal Institute for Astrophysics and Machine Learning, Montreal QC, Canada}
\affiliation{Mila - Quebec AI Institute, Montreal QC, Canada}
\affiliation{Department of Physics, University of Montréal, Montréal QC, Canada}

\author[0000-0002-8669-5733]{Yashar Hezaveh}
\affiliation{Ciela Institute, Montreal Institute for Astrophysics and Machine Learning, Montreal QC, Canada}
\affiliation{Mila - Quebec AI Institute, Montreal QC, Canada}
\affiliation{Department of Physics, University of Montréal, Montréal QC, Canada}
\affiliation{Center for Computational Astrophysics, Flatiron Institute, NY, USA}
\affiliation{Perimeter Institute for Theoretical Physics, Waterloo, Canada}
\affiliation{Trottier Space Institute, McGill University, Montréal, Canada}

\author[0000-0003-3544-3939]{Laurence Perreault-Levasseur}
\affiliation{Ciela Institute, Montreal Institute for Astrophysics and Machine Learning, Montreal QC, Canada}
\affiliation{Mila - Quebec AI Institute, Montreal QC, Canada}
\affiliation{Department of Physics, University of Montréal, Montréal QC, Canada}
\affiliation{Center for Computational Astrophysics, Flatiron Institute, New York, USA}
\affiliation{Perimeter Institute for Theoretical Physics, Waterloo, Canada}
\affiliation{Trottier Space Institute, McGill University, Montréal, Canada}

\author[0000-0002-5116-7287]{Daniel Gilman}
\affiliation{University of Chicago, Chicago IL, USA}

\author{LSST Dark Energy Science Collaboration}

\begin{abstract}
Strong gravitational lensing offers a unique probe of dark matter (DM) on sub-galactic scales, where the abundance and distribution of low-mass halos are highly sensitive to the underlying properties of DM particles. 
In this work, we forecast LSST’s sensitivity to DM substructure in galaxy-galaxy strong lenses using simulated samples and neural ratio estimators (NREs). Our simulations include both subhalos within the main deflector and line-of-sight (LOS) halos, with halo masses down to $\sim 10^7 M_\odot$ under the expected LSST ten-year survey imaging quality. 
We show that the constraining power on halo mass function (HMF) parameters improves significantly with sample size. Analyses based on a few hundred lenses yield broad posteriors comparable with other probes like the Ly-$\alpha$ forest. By contrast, when combining 2\,500 lenses, $\approx 74\%$ and $\approx 36\%$ of the prior volume considered can be excluded at the $3\sigma$ and $5\sigma$ levels respectively, enabling statistically significant exclusions of non-$\Lambda$CDM scenarios.
We further demonstrate that the sensitivity arises not only from the high-mass end of the HMF but also from low-mass halos: masking halos below $\log (m_{\rm halo}/M_\odot) \leq 7.5$ induces a measurable shift in the inferred posteriors. Finally, we find that LOS halos contribute significantly to the constraining power, with increasing importance of LOS halos at higher redshifts.
While this analysis assumes perfect knowledge of the data-generating process and cannot be directly applied to data analysis, it quantifies constraints achievable with LSST alone and motivates the development of robust inference methods for real survey data.
\end{abstract}


\section{Introduction} \label{sec:intro}
The Vera C. Rubin Observatory's \textit{Legacy Survey of Space and Time} (LSST) will transform our understanding of the Universe's fundamental constituents. Its 10-year mission will map the southern sky in six optical bands, achieving a coadded 5-$\sigma$ depth of $r \sim27.5$ and a median r-band seeing of $\sim0.7''$ \citep{Collett_2015}.

One of LSST's primary science goals is the study of dark matter. Although it constitutes roughly 80\% of the Universe’s total matter content, its nature remains unknown \citep[e.g.,][]{WMAP_2013, Planck_2020}. The presence of dark matter is inferred from gravitational phenomena across cosmic scales, from galaxy rotation curves to the acoustic peaks in the Cosmic Microwave Background. However, it has not been detected via non-gravitational interactions, leaving its fundamental properties elusive.

Different dark matter models predict distinct clustering behaviors, especially on sub-galactic scales. These differences manifest in the populations of both subhalos (halos populating galaxies) and line-of-sight (LOS) halos. 
On these small scales, the dark matter distribution, quantified by the halo mass function (HMF), is highly sensitive to the particle nature of dark matter, making the HMF a powerful model discriminator \citep[e.g.,][]{Ferreira_2021_DM}. For example, in warm dark matter (WDM) models, the dark matter particles have thermal velocities that suppress the formation of low-mass halos below a free-streaming scale \citep[e.g.,][]{Colin_2000_WDM, Gilman_2020_HMF, Loudas_2022_WDM}. In contrast, cold dark matter (CDM) models predict hierarchical clustering down to very small scales, resulting in a continuously rising low-mass HMF and abundant substructure on these scales.

Strong gravitational lensing offers a unique method to probe the matter distribution on these small scales, for both subhalos and LOS halos. Unlike techniques that rely on luminous tracers, lensing is sensitive to all matter - luminous or dark - making it a powerful observational tool for constraining the halo mass function. Traditional analyses infer the presence of individual halos by assessing whether introducing localized perturbers to a smooth lens model yields a statistically significant improvement in the fit to the observed lensed images \citep[e.g.,][]{Vegetti_2010, Yashar_2016}. 

To move beyond those individual detections, inferring dark matter properties from strong lensing observations is fundamentally a hierarchical Bayesian task. It requires marginalizing over individual source and lens parameters, as well as specific subhalo realizations for every analyzed system, to obtain population-level constraints on the mass function. While this target posterior is low-dimensional (e.g., one or two parameters), the process involves marginalizing over high-dimensional spaces spanning thousands of dimensions, rendering traditional sampling methods computationally prohibitive. Recent progress in machine learning methodology, more specifically simulation-based inference methods, has been demonstrated to have high potential for population-level inference of the HMF by combining data across ensembles of lensing systems \citep[e.g.,][]{brehmer2019mining, Brehmer_Sidd_2019_NRE, Coogan_2022, Wagner_C2023, Wagner_C2024, Zhang_2024, OOD_paper_24, Sidney_2025_01, Sydney_2025, Padma_2025, Pho_2025}.

One of LSST's most exciting prospects is its anticipated discovery of approximately $10^5$ strong gravitational lenses \citep[e.g.,][]{LSST_Science_Book_2009, Serjeant14, Collett_2015, Ivezic_LSST_dataproducts_2019}.
This dramatic increase over existing samples opens new opportunities for statistical studies of dark matter through perturbations induced in strongly lensed images. 
The question we aim to answer here is that of the sensitivity of this expected sample to the properties of the underlying population-level HMF, assuming the full 10-year survey is completed. For this forecast, we leverage Neural Ratio Estimators (NREs), which are a specific class of simulation-based inference (SBI) methods. NREs are particularly well-suited for this task, as they allow amortization of the inference process as well as implicit marginalization over very high-dimensional spaces to infer low-dimensional variables. To do so, NREs directly learn the likelihood-to-evidence ratio by making use of a sufficiently realistic simulation pipeline, enabling scalable inference \citep[e.g.,][]{brehmer2019mining, Brehmer_Sidd_2019_NRE, Coogan_2022, Jarugula_2024}. 
While NREs typically require knowledge of the true data-generating distribution - a limitation in real-world analysis applications \citep[e.g.,][]{OOD_paper_24} - we can assume such knowledge for this forecasting study.

In this work, we present a forecasting analysis of LSST's sensitivity to dark matter substructure via strong gravitational lensing. We focus exclusively on LSST data and do not assume auxiliary observations from other surveys such as \textit{Euclid} or \textit{Roman}. By simulating realistic lensing observations and applying NREs, we also explore which halo populations dominate the constraining power: specifically, whether this power is driven by low-mass/high-abundance or high-mass/low-abundance halos, and the relative contributions of subhalos versus line-of-sight halos.

This paper is organized as follows: Section~\ref{sec:data_generation} details the data generation process, including relevant LSST observational characteristics and the lensing simulations used. In Section~\ref{sec:methods}, we introduce the NRE methodology and the data processing pipeline. Our forecast results are presented in Section~\ref{sec:results}, where we also address which halo populations dominate the sensitivity to the dark matter mass function. Finally, we discuss broader implications for dark matter research and summarize our key findings in Section~\ref{sec:discussion}.

\section{Data Generation}  \label{sec:data_generation}
\subsection{Strong Lensing}
Strong gravitational lensing occurs when the light from a distant background source ($S$) is deflected and distorted by a foreground mass distribution, typically a massive galaxy or galaxy cluster, known as the main deflector. This gravitational effect can produce characteristic features such as multiple images, arcs, and rings, depending on the alignment between the source, lens, and observer. The total mass density of the lens is generally modeled as two distinct components: a smooth, large-scale density profile representing the main deflector and small-scale perturbations contributed by low-mass dark matter subhalos embedded within the deflector's host halo. Additionally, dark matter halos along the line-of-sight (LOS) between the source and the observer further perturb the lensed source.

The number and exact distribution of both subhalos and LOS halos are not known a priori, which makes likelihood inference of individual halo properties a transdimensional problem. Subhalo populations depend on the properties of the main deflector (such as its mass), whereas LOS halos follow a statistical distribution predicted by the $\Lambda$CDM model and are not directly tied to the lens galaxy. The LOS distribution is typically approximated by the Sheth–Tormen mass function \citep[][]{Sheth-Tormen_2001}, which approximately follows a power law. Consequently, both halo populations (subhalos and LOS halos) are modeled using mass functions parameterized as power laws, each defined by an amplitude and a slope.

To allow for deviations from $\Lambda$CDM predictions, we introduce two scaling parameters: $A$, which modifies the amplitude of the mass function, and $\gamma$, which modifies its slope. 
For our sensitivity forecasts, we are interested in these two parameters, which govern the subhalo and LOS halo mass functions relative to the $\Lambda$CDM prediction, summarized collectively as the parameter set $\vartheta = (A,\gamma)$. The observed lensed images ${D_i}$ are determined not only by $\vartheta$ but also by a set of nuisance parameters $\theta$, which include properties such as the Einstein radius, ellipticity, source parameters, and the positions and masses of the individual dark matter halos.

To simulate the lenses, we utilize \href{https://github.com/Ciela-Institute/caustics}{\texttt{caustics}}\footnote{\url{https://github.com/Ciela-Institute/caustics}} \citep[][]{Caustics_2024}. 
We model the light of the background source with an ensemble of 5 to 50 Sérsic profiles \citep[][]{Sersic_1963} to represent sources of varying morphological complexity. The main deflector is modeled as a singular isothermal ellipsoid (SIE), for which the normalized surface mass density, $\kappa$, is defined as
\begin{equation}
    \kappa(x, y) = \frac{1}{2} \left(\frac{\theta_{\rm E}}{\sqrt{q x^2 + y^2/q}} \right)
    \label{eq:EPL_kappa}
\end{equation}
where $\theta_{\rm E}$ is the Einstein radius and $q$ is the axis ratio of the mass profile. The coordinates $(x,y)$ define a Cartesian coordinate system aligned with the major and minor axes \citep[][]{Tessore_EPL}.

The Einstein radius $\theta_{\rm E}$ depends on the mass enclosed within it, $M(\theta_{\rm E})$:
\begin{equation}
    \theta_{\rm E} = \sqrt{\frac{4GM(\theta_{\rm E})}{c^2} \frac{D_{\rm ls}}{D_{\rm l} D_{\rm s}}}
\end{equation}
and defines the typical image separation scale.

Individual dark matter halos — both subhalos and LOS halos — are modeled using truncated Navarro-Frenk-White (tNFW) profiles with truncation radius $r_t$ \citep[see,][]{NFW_profile, tNFW_2009}: 
\begin{equation}
    \rho_{\rm tNFW}(r) = \frac{M_0}{4 \pi r (r + r_s)^2} \left( \frac{r_t^2}{r^2 + r^2_t} \right)^n
\end{equation}

Using \texttt{caustics}, we simulate LOS halos and subhalos in a way that their total mean convergence is zero. This is achieved by subtracting the global mean convergence of the halo population from the convergence field of each individual halo, effectively redistributing the mass to ensure the average background density remains unchanged. This means that the Einstein radius and effective slope of the smooth component of the main lens convergence will not be affected by the addition of a population of subhalos.

The mass for each halo is drawn from its respective mass function. The power-law mass function for subhalos associated with the main deflector is given by:
\begin{equation}
    \frac{\text{d}n}{\text{d log } m_{\rm halo}} = \alpha \cdot M_{\rm host} \cdot m_{\rm halo}^{\beta}
    \label{eq:dm_mf}
\end{equation}
where $\alpha$ is the normalization constant, $M_{\rm host}$ is the mass of the main deflector, and $m_{\rm halo}$ is the subhalo mass.

To quantify and normalize the abundance of dark matter in subhalos, we introduce the parameter $f_{\rm sub}$. This parameter defines the ratio of the total mass contained in dark matter subhalos to the mass of the main deflector (as defined by the SIE profile) and sets the normalization $\alpha$:
\begin{equation}
\begin{split}
    f_{\rm sub} &= \alpha \int \text{d}m_{\rm halo} m_{\rm halo}^{\beta+1} \\
    &= \frac{\int \text{d}m_{\rm halo} m_{\rm halo} \frac{\text{d}n}{\text{d}m_{\rm halo}}}{M_{\rm host}}
\end{split}
\end{equation}
This formulation enables the efficient sampling of dark matter subhalos across different dark matter mass functions. In this work, we assume that the parameters $(f_{\rm sub}, \beta)$ fully define the dark matter subhalo population.
The $\Lambda$CDM prediction for this power-law approximation is $f_{\rm sub,CDM}\approx0.05$ and $\beta_{\rm CDM} \approx -0.9$ \citep[e.g.,][]{Diemand_2007_ViaLactea, Kuhlen_2007_ViaLactea, Springel_2008, Hiroshima_2018}.

LOS halo masses are drawn from a simplified Sheth–Tormen HMF at $z=0$ \citep[][]{Sheth-Tormen_2001}. This HMF is fitted with a power law between $\log_{10} M = 7$ and $10$, taking the form:
\begin{equation}
    \frac{\text{d}n}{\text{d log } m_{\rm los} \text{d}V} = A_{\rm CDM} \cdot m_{\rm los}^{\gamma_{\rm CDM} +1}
    \label{eq:LOS_HMF}
\end{equation}
where $V$ represents the volume of a redshift bin. The $\Lambda$CDM best-fit parameters at $z=0$ yield $A_{\rm CDM} = 9.35$ and $\gamma_{\rm CDM} = -2.03$.

To explore alternative DM models, we define a two-parameter extension to $\Lambda$CDM that simultaneously modifies the subhalo and line-of-sight (LOS) populations. We introduce an amplitude parameter, $A$, which rescales the normalizations ($f_{\rm sub,CDM}$ and $A_{\rm CDM}$), and a slope parameter, $\gamma$, which rescales the power-law slopes ($\beta_{\rm CDM}$ and $\gamma_{\rm CDM}$).
This allows us to explore empirical deviations from the $\Lambda$CDM model in a model-agnostic way, where the parameters $(A, \gamma) = (1,1)$ correspond to $\Lambda$CDM predictions. 
When varying the slope $\gamma$, we normalize the amplitude of the power-law to have the same integrated mass in dark matter halos as for the corresponding CDM slope.
In order to investigate data-driven constraints, we choose not to use priors informed by specific DM alternative models (however, note that, as demonstrated in \citet{Toomey_Desi_2025}, such choices could significantly affect the constraints obtained). We also note that the parametrization chosen is not optimal for some classes of alternative models, such as fuzzy or warm DM \citep[e.g.,][]{FuzzyCDM_2000, WDM_2001}. 
The chosen parametrization can imitate certain settings of alternative dark matter models, for example, specific choices of slope and amplitude factors can correspond to different WDM models. Whilst no explicit comparison with other DM models is possible, this work shows that LSST data does have constraining power to gain DM particle properties.

\begin{figure}[t]
    \centering
    \includegraphics[width=0.95\linewidth]{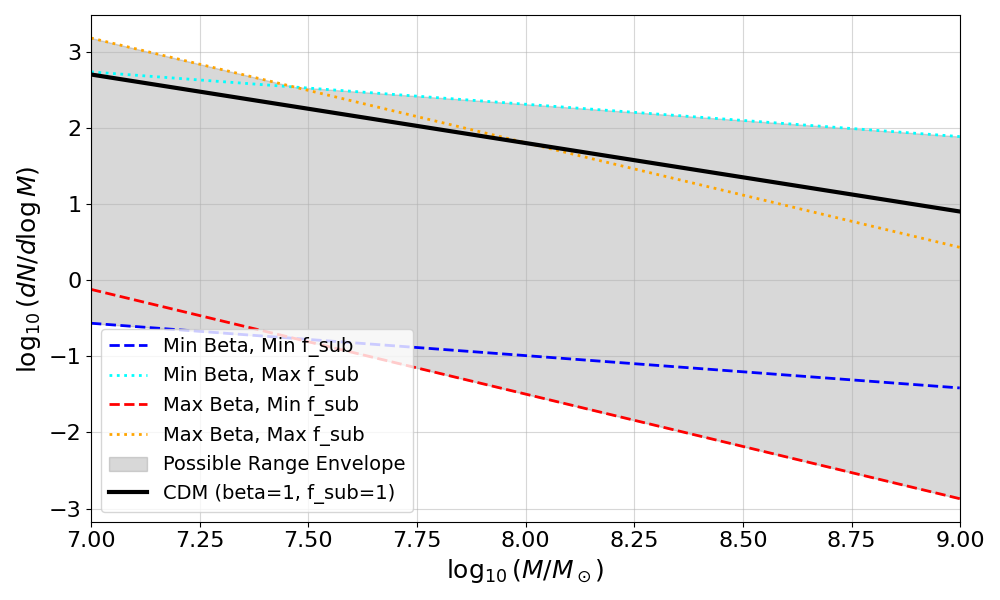}
    \caption{Range of possible DM HMFs with this parameterization with respect to the CDM parameterization, shown by the black solid line. The gray covers all the possibilities, with the minimal $A$ being limited to $A>10^{-3}$ for illustration purposes. $A=0$ corresponds to no subhalos for any power-law slope $\gamma$.}
    \label{fig:powerlaw_poss}
\end{figure}

Figure~\ref{fig:powerlaw_poss} illustrates the range of possible DM HMFs with this parameterization with respect to the CDM parameterization, shown by the black solid line. The gray area covers all the possibilities, with the minimal $A$ being limited to $A>10^{-3}$ for illustration purposes. $A=0$ corresponds to no subhalos for any power-law slope $\gamma$.

\begin{figure}[t]
    \centering
    \includegraphics[width=0.95\linewidth]{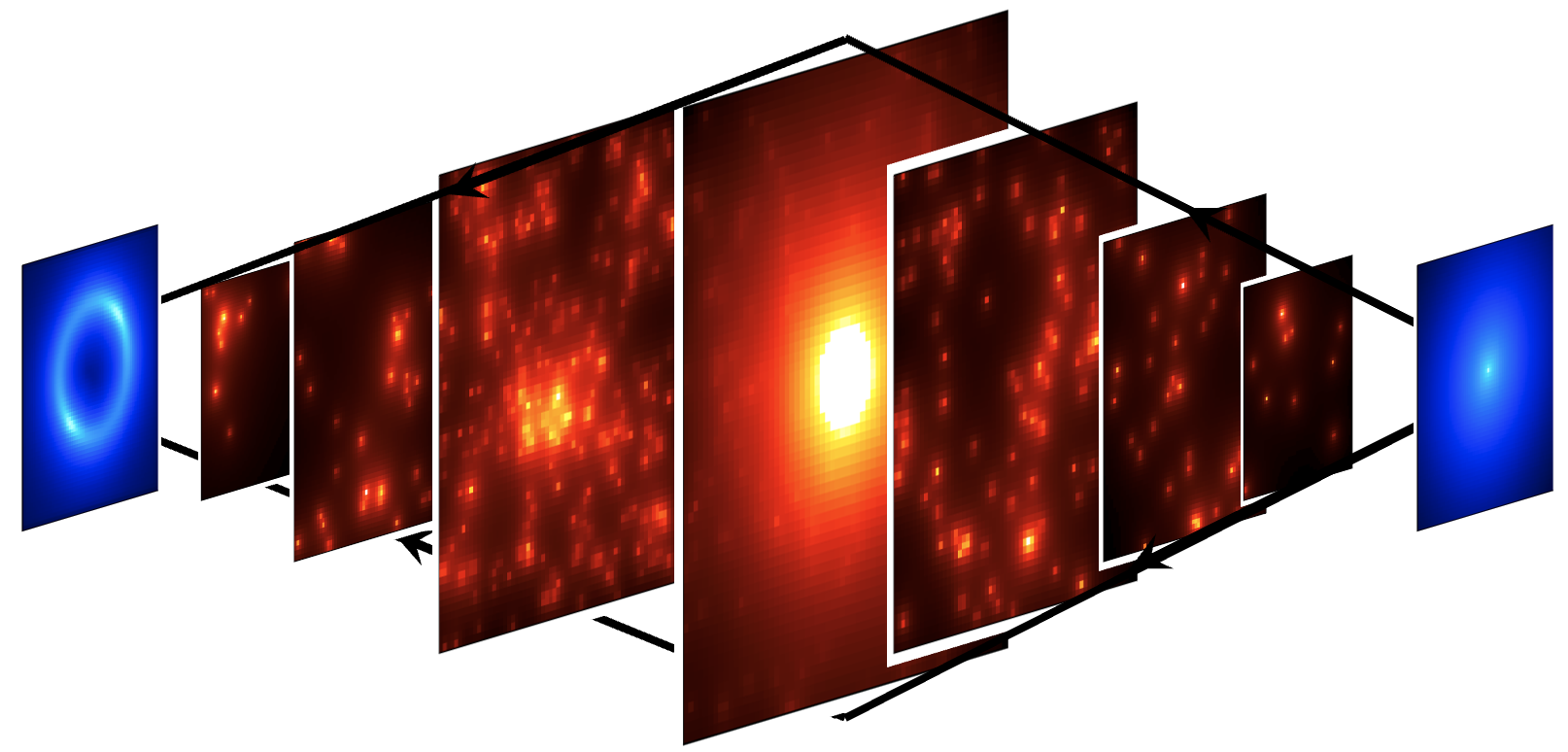}
    \caption{Illustration of the convergence maps to raytrace through in multi plane lensing at different redshifts, with the main deflector in the middle.}
    \label{fig:multiplane}
\end{figure}

We divide the LOS into 10 equally spaced redshift bins and compute the comoving volume within each bin to determine the halo counts. Since the higher redshift bins contain a larger volume, they naturally host a greater number of halos. We then ray-trace through all LOS planes and the main deflector plane using the dominant-lens approximation.
Figure~\ref{fig:multiplane} illustrates the convergence maps to raytrace through in multi-plane lensing at different redshifts, with the main deflector in the middle.

The standard multiplane lens equation is:
\begin{equation}
    \vec{\beta} = \vec{\theta} - \sum _{i=1} ^{N} \frac{D_{\rm is}}{D_{\rm s}} \vec{\widetilde\alpha}_i (\vec{x_i})
    \label{eq:multiplanerecursive}
\end{equation}
where $\vec{\beta}$ is the unlensed source position, $\vec{\theta}$ the image position, $\vec{\widetilde\alpha}_i$ the deflection angle in plane $i$, and $D_{\rm is}$ and $D_{\rm s}$ the angular diameter distances. $N$ is the number of planes we raytrace through, which, in our case, corresponds to the number of redshift bins plus the plane of the main lens, $N=11$.

Because the deflection of a ray at any given plane depends recursively on all of the previous deflections it incurred, solving the recursive lens equation is computationally expensive, particularly when incorporating many lens planes containing numerous deflectors. To mitigate this complexity, we employ the dominant-lens approximation, which assumes a single primary deflector and treats all other lens planes as perturbative.

Following the notation of \citet{Fleury_2021} and expanding to first order in the deflection angles  $\vec{\alpha}_i$ of the line-of-sight halos, the multiplane lens equation~\ref{eq:multiplanerecursive} can be simplified to:
\begin{equation}
\begin{split}
    \vec{\beta} &= \vec{\theta} - \vec{\alpha}(\vec{\theta}) \\
    \vec{\alpha}(\vec{\theta}) &= \underbrace{ \frac{D_{\rm ls}}{D_{\rm s}} \vec{\alpha}_l(\vec{\theta}) }_{\text{dominant}} + 
    \underbrace{ \sum _{i<l} \frac{D_{\rm is}}{D_{\rm s}} \vec{\alpha}_i(\vec{\theta}) }_{\text{foreground}} \\
    &+ 
    \underbrace{ \sum_{i>l} \frac{D_{\rm is}}{D_{\rm s}} \vec{\alpha}_i \left[ \vec{\theta} - \frac{D_{\rm li}}{D_{\rm i}}\vec{\alpha}_l(\vec{\theta}) \right] }_{\text{background}} 
    + \mathcal{O}(\vec{\alpha}_i^2)
    \label{eq:dominant_lens}
\end{split}
\end{equation}
where the background term depends recursively on the dominant lens, with $\vec{\alpha}_l$ the deflection angle of the dominant lens. Notably, the background term remains coupled to the dominant lens deflection since it is treated non perturbatively, and $\mathcal{O}(\vec{\alpha}_i^2)$ denotes neglected higher-order perturber terms.

We do not model the lens galaxy light, effectively assuming perfect subtraction of the foreground light component. While not achievable in practice, this assumption is useful for establishing the intrinsic limits of LSST-quality data.

Table~\ref{tab:parameter_distribution} lists the parameter distributions used in training and testing. We simulate systems for lens redshifts from $z_{\rm l}$ ranging from $0.2$ to $1.0$ in steps of $\Delta z_{\rm l} =0.2$, and source redshifts $z_{\rm s}$ from $0.25$ to $4.0$ in steps of $\Delta z_{\rm s} =0.25$. This results in 70 unique redshift combinations.

\begin{table}[t]
    \centering
    \caption{Parameter distributions used in the simulation of the main deflector, source light, and DM parameters for training the NRE. Distributions are denoted as $\mathcal{N}(\mu, \sigma)$ for a normal distribution and $\mathcal{U}(a, b)$ for a uniform distribution.}
    \renewcommand{\arraystretch}{1.2}
    \begin{tabular}{l l}
    \hline \hline
    \textbf{Parameter} & \textbf{Distribution} \\ 
    \hline
    \multicolumn{2}{l}{\textbf{Lens galaxy}} \\ \hline
    Einstein radius $\theta_{\rm E}$ & $\mathcal{U}(1.0, 1.5)$ \\
    Axis ratio $q_{\rm SIE}$ & $\mathcal{U}(0.5, 0.99)$ \\ 
    Orientation angle $\phi_{\rm SIE}$ & $\mathcal{U}(0, \pi)$ \\ 
    Lens center $(\hat{x}_{\rm SIE}, \hat{y}_{\rm SIE})$ & Fixed at $(0, 0)$ \\
    \hline
    \multicolumn{2}{l}{\textbf{Parameters of interest}} \\ \hline
    DM amplitude factor $A$ & $\mathcal{U}(0.0, 2.0)$ \\
    DM mass slope factor $\gamma$ & $\mathcal{U}(0.75, 1.25)$ \\
    \hline
    \multicolumn{2}{l}{\textbf{Source light}} \\ \hline
    Number of Sérsic components $N$ & $\mathcal{U}(5, 50)$ \\
    Magnitude $mag_{\rm source}$ & $\mathcal{N}(23.5, 0.1)$ \\
    Sérsic index $n_{\rm S\acute{e}rsic}$ & $\mathcal{N}(2.5, 0.5) \geq 0.8$ \\
    Axis ratio $q_{\rm S\acute{e}rsic}$ & $\mathcal{U}(0.5, 0.99)$ \\
    Orientation angle $\phi_{\rm S\acute{e}rsic}$ & $\mathcal{U}(0, \pi)$ \\
    Sérsic radius $R_{\rm S\acute{e}rsic}$ & $\mathcal{N}(0.5, 0.3) \geq 0.05$ \\
    Source center $(\hat{x}_{\rm source}, \hat{y}_{\rm source})$ & $\mathcal{N}(0.0, 0.1)$ \\
    \hline \hline
    \end{tabular}
    \label{tab:parameter_distribution}
\end{table}

\subsection{LSST data specifications}
The data specifications used for training and testing the NRE are based on the expected LSST data quality in the r-band after the completion of the full ten-year survey \citep[see][]{LSST_numbers_2019}.  
In our analysis, we exclusively use the r-band specifications corresponding to the coadded 10-year dataset. The relevant parameters are listed in Table~\ref{tab:lsst_specs}, with values adapted from \citet{LSST_numbers_2019} and \citet{SMTN_002}.

The instrumental noise is approximated by the following formula:
\begin{equation}
    \sigma_{\rm instr}^2 = (rN^2 + (dC * t_{\rm exp})) * n_{\rm exp}
\end{equation}
where the readout noise is $rN = 8.8 \, e^{-}/{\rm pixel \, exposure}$, and the upper limit on the dark current is $dC = 0.2 \, e^{-}/{\rm s \, / pixel}$.  
The observing strategy includes two back-to-back exposures per visit (i.e., $n_{\rm exp} = 2$ exposures per visit), with each individual exposure having a duration of $t_{\rm exp} = 15\,{\rm s}$.  
Following Figure 18 in \citet{Ivezic_LSST_dataproducts_2019}, we assume a total of 200 visits per lens system.

The photometric zero point, sky background magnitude, pixel scale, and median seeing in the r-band are listed in Table~\ref{tab:lsst_specs}.  
The source magnitudes used in our simulations (Table~\ref{tab:parameter_distribution}) are fainter than those reported by \citet{Collett_2015}, who focused on g-band magnitudes. In addition, brighter lenses are generally easier to detect, which justifies our choice of brighter r-band magnitudes for forecasting purposes.

\begin{table}[t]
    \centering
    \caption{Key LSST r-band specifications used for lensing simulations. Values are based on a 10-year coadded survey.}
    \renewcommand{\arraystretch}{1.2}
    \begin{tabular}{l c l}
    \hline \hline
    \textbf{Specification} & \textbf{Value} & \textbf{Unit} \\
    \hline
    Survey duration & 10 & years \\
    Sky coverage & $\sim$18,000 & deg$^2$ \\
    Median seeing (r-band) & 0.83 & arcsec \\
    Photometric zero point (r-band) & 28.36 & mag \\
    Sky brightness (r-band) & 21.2 & mag/arcsec$^2$ \\
    Cadence & $\sim$200 & visits/field \\
    Exposure time per visit & $2 \times 15$ & s \\
    Pixel scale & 0.2 & arcsec/pixel \\
    \hline \hline
    \end{tabular}
    \label{tab:lsst_specs}
\end{table}

To model the redshift distribution of the lenses and sources, we approximate the expected distribution reported in Figure 1 of \citet{Collett_2015}.  
These predictions are consistent with recent JWST observations \citep[][]{Hogg_2025_JWST_forecasting}, supporting their validity as forecasts for LSST.

\begin{figure}[t]
    \centering
    \includegraphics[width=.99\linewidth]{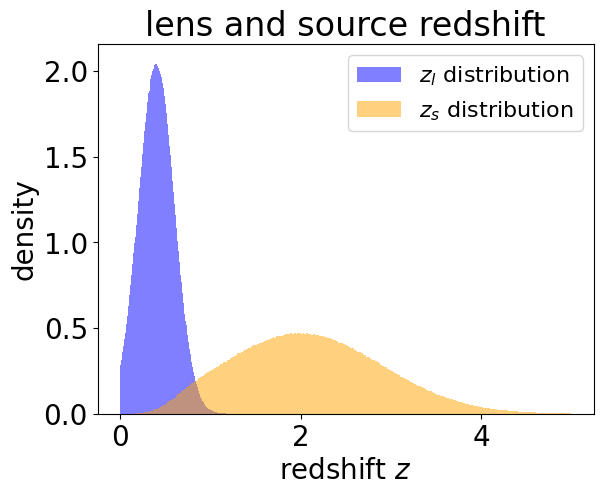}
    \caption{Redshift distributions of lenses and sources used in the simulation. Each sampled source redshift is constrained to be greater than its corresponding lens redshift.}
    \label{fig:redshift_dists}
\end{figure}

Figure~\ref{fig:redshift_dists} shows the redshift distributions from which we sample. We strictly enforce the physical condition that the source redshift ($z_{\rm s}$) must be greater than the corresponding lens redshift ($z_{\rm l}$).

\begin{figure}[t]
    \centering
    \includegraphics[width=.99\linewidth]{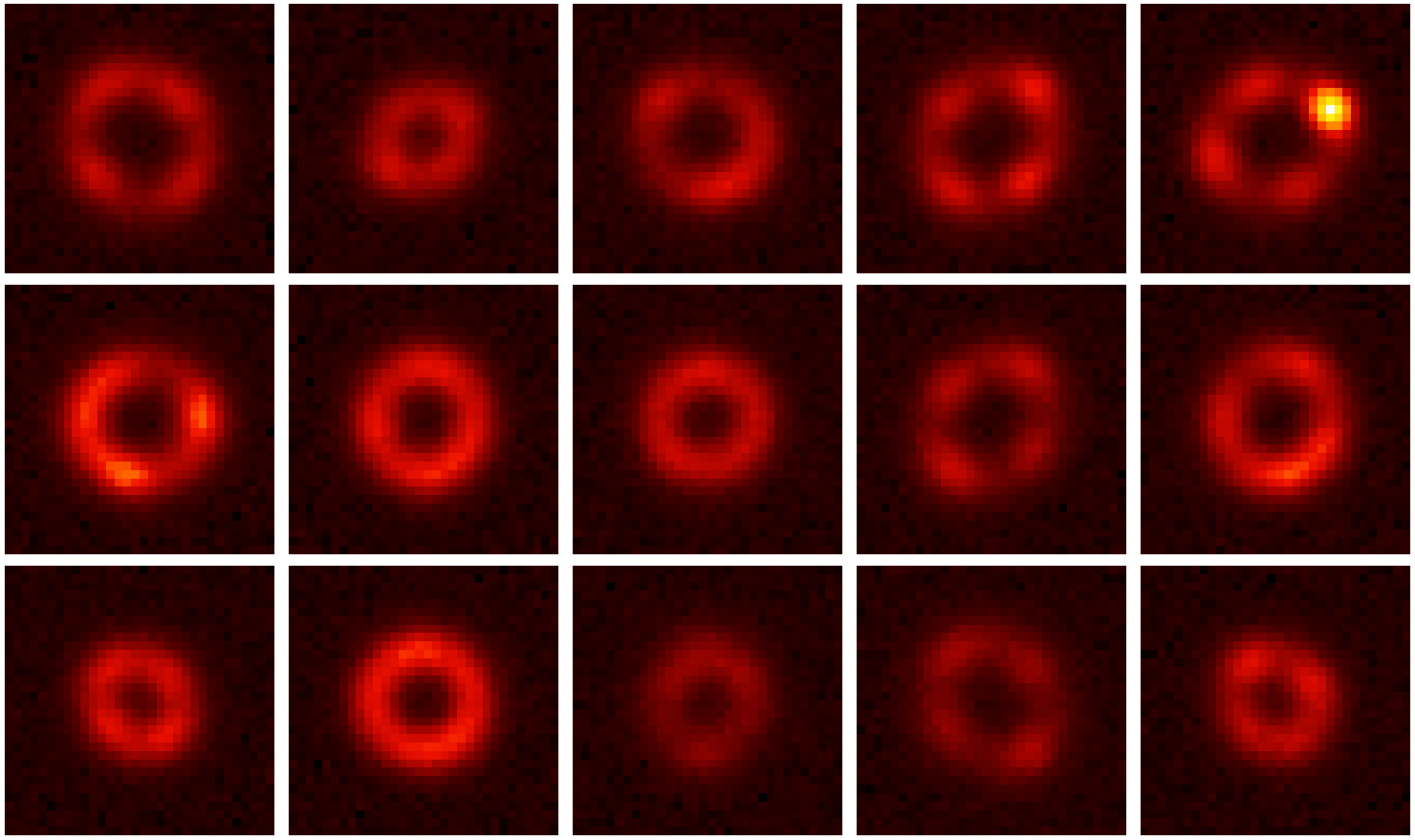}
    \caption{Fifteen example lensing images generated from simulated LSST data. The set includes quadruply imaged systems (quads), doubles, and Einstein rings.}
    \label{fig:images}
\end{figure}

Figure~\ref{fig:images} presents 15 example lens images generated using the described LSST and simulation specifications. The examples include quads, doubles, and Einstein rings, highlighting the morphological diversity captured in our dataset.

\section{Methods} \label{sec:methods}

\subsection{Bayesian Inference in Strong Lensing} 
The foreground mass density is the sum of two components: a smooth profile for the main deflector and small-scale fluctuations from DM subhalos. The smooth component is represented by the analytic Singular Isothermal Ellipsoid (SIE) profile, with $L$ denoting the parameters of this profile. Individual DM halos $m_i$ — comprising both subhalos and LOS halos — are modeled using truncated NFW (tNFW) profiles \citep[see,][]{NFW_profile, tNFW_2009}.

Given the large number of these halos, we use $H=\{m_1,m_2,...\}$ to denote the set of parameters for the entire halo population (e.g., redshifts, positions, masses, and truncation radii). Because the number of halos is inherently unknown, the dimensionality of $H$ varies across different lens systems, making the problem of inferring their properties a transdimensional problem.
 
The parameters of individual halos, $m_i$ (specifically, their masses $m_{h,i}$ and positions $\psi_i$ in the image plane), are drawn from a population-level distribution governed by the parameters of interest, $\vartheta$ (the normalization and slope of the DM mass functions, see Section~\ref{sec:data_generation}). Our goal is to determine the posterior distribution of these population parameters, marginalized over all nuisance parameters, $S$, $L$, and $H$. For simplicity, we collectively denote all nuisance parameters as $\theta = \{ S,L,H \}$ and the observed data as $D$. 

The observations are, in our case, images of strong lenses that are not seen during training. A set of images is denoted by $\{D_i\}$. They depend on the nuisance parameters $\theta$, which themselves depend on the population-level parameters $\vartheta$.
Thus, the likelihood $p(D_i|\vartheta)$ is obtained by marginalizing the joint likelihood $p(D_i,\theta|\vartheta)$ over all intermediate parameters $\theta$:
\begin{equation}
    p(D_i|\vartheta) = \int \text{d}\theta \, p(D_i,\theta|\vartheta) .
    \label{eq:goal_likelihood}
\end{equation}
Both the likelihood $p(D_i|\vartheta)$ and the posterior $p(\vartheta|\{D_i\})$ are intractable to compute analytically. This intractability arises from the high-dimensional integral required for marginalization, as the joint likelihood $p(D_i,\theta|\vartheta)$ can be expressed as
\begin{equation}
    \begin{split}
        p(D,&\theta|\vartheta) = p_{\rm lens}(L) \\ 
        &\quad \times \text{Pois}(n_{h}|\bar{n}_{h}(\vartheta)) \prod_{i=1}^{n_h}\left[p_{\rm mass}(m_{h,i}|\vartheta) p(\psi_i)\right] \\
        &\quad \times p_{\rm obs}(D|\theta) 
    \end{split}
    \label{eq:simplify_likelihood}
\end{equation}
where $p_{\rm lens}(L)$ is the prior distribution of lens parameters, $\bar{n}_h(\vartheta)$ is the expected number of halos given $\vartheta$, and $n_h$ is the realized number of halos in a specific simulation. 

By combining independent observations $\{D_i\}$, we can estimate a population-level posterior for the parameters of interest $\vartheta$. We use Bayes’ theorem to obtain the posterior as:
\begin{equation}
\begin{split}
    p(\vartheta|\{D_i\}) &= \frac{p(\vartheta) \prod_i p(D_i|\vartheta)}{\int \text{d}\vartheta' p(\vartheta') \prod_i p(D_i|\vartheta')} \\
    &= p(\vartheta) \left[ \int \text{d}\vartheta' \, p(\vartheta') \prod_i \frac{p(D_i|\vartheta')}{p(D_i|\vartheta)} \right]^{-1},
    \label{eq:posterior}
\end{split}
\end{equation}
where $p(\vartheta)$ represents the prior probability distribution on the population-level parameters.
To obtain a posterior distribution of the parameters of interest, we create a grid of possible parameters of interest $\vartheta$ and evaluate the posterior for a given set of images $\{D_i\}$ at each possible $\vartheta$.

\subsection{Neural Ratio Estimator}
For forecasting, we train and evaluate 70 Neural Ratio Estimators (NREs).  
We trained one for each redshift combination individually because a single network with a condition on the redshift yielded less constraining posterior results.
We utilize Neural Ratio Estimators (NREs) for this forecasting analysis. Although NREs can produce biased results if the underlying data-generating process is misspecified \citep[][]{OOD_paper_24}, they are exceptionally well-suited for forecasting applications where the true data-generating process is known exactly.

Rather than directly learning the likelihood $p(D_i|\vartheta)$ or the posterior $p(\vartheta|D_i)$, the training objective of NREs is the likelihood-to-evidence ratio $r(D, \theta|\vartheta)$:
\begin{equation}
    r(D,\theta|\vartheta) = \frac{p(D,\theta|\vartheta)}{p_{\rm ref}(D, \theta)}\, ,
    \label{eq:ratio}
\end{equation}
where the reference distribution, $p_{\rm ref}(D, \theta)$, is the marginal of observing $(D,\theta)$ under any possible DM parameterization $\vartheta$, defined as
\begin{equation}
    p_{\rm ref}(D,\theta) = \int \text{d}\vartheta' \, p(\vartheta')p(D, \theta|\vartheta') \,.
\end{equation}
Here, $p(D, \theta|\vartheta')$ is the likelihood of an observation $D$ given a specific set of DM parameters $\vartheta'$, and $p(\vartheta')$ is the prior distribution for the DM HMF parameters used in the training data generation.

Predicting the likelihood-to-evidence ratio $r(D,\theta|\vartheta)$, rather than the likelihood or posterior directly, enables efficient training. When substituting Equation~\ref{eq:simplify_likelihood}, the ratio simplifies significantly to
\begin{equation}
        r(D,\theta|\vartheta) = 
        \text{Pois}(n_{h}|\bar{n}_{h}(\vartheta)) \prod_{i=1}^{n_h}\left[p_{\rm mass}(m_{h,i}|\vartheta) p(\psi_i) \right] \, ,
\end{equation}
since all other terms cancel when forming the ratio, as they do not depend on the parameters of interest $\vartheta$. This resulting expression becomes analytic if the underlying halo mass function is also analytic \citep[see,][]{Cranmer_2015,brehmer2018constraining, Brehmer_Sidd_2019_NRE, brehmer2019mining, Coogan_2022, Jarugula_2024}.  
By training the network on simulated images $D$ generated with different parameters $(\theta, \vartheta)$ and their corresponding likelihood-to-evidence ratio $r(D,\theta|\vartheta)$, the NRE learns to marginalize over nuisance parameters $\theta$.

The trained NRE predicts the marginal likelihood-to-evidence ratio $r(D|\vartheta)$, which allows us to rewrite Equation~\ref{eq:posterior} using:
\begin{equation}
    \frac{p(D_i|\vartheta')}{p(D_i|\vartheta)} = \frac{p(D_i|\vartheta')}{p_{\rm ref}(D_i)}/\frac{p(D_i|\vartheta)}{p_{\rm ref}(D_i)} = \frac{r(D_i|\vartheta')}{r(D_i|\vartheta)}
\end{equation}
Hence, the population-wide posterior becomes
\begin{equation}
    p(\vartheta|\{D_i\}) = p(\vartheta) \left[\int \text{d}\vartheta' p(\vartheta') \prod_i \frac{r(D_i|\vartheta')}{r(D_i|\vartheta)} \right]^{-1}
\end{equation}
Crucially, the remaining integral is only over the parameter space of interest $\vartheta$ (encompassing the normalization and slope factors of the HMF), allowing for direct calculation of the posterior.
To obtain the posterior distribution, we create a grid of all possible parameters of interest $\vartheta$ and evaluate the NRE for a given set of images $\{D_i\}$ at each possible $\vartheta$.

We train each of the 70 different NRE on 1 million lens images. For calibration, we generate for each single NRE 15\,000 strong lenses for the reference distribution, and 5\,000 systems for each possible considered bin of HMF parameter combination.

Details of the training procedure and calibration of the NRE (e.g., optimizer, learning rate, loss function, etc.) can be found in Appendix~\ref{app:NRE_training}.

In the limit of infinite data and perfect training, it can be shown that the NRE converges to an estimator that recovers the exact marginal likelihood-to-evidence ratio \citep[e.g.,][]{Cranmer_2015, brehmer2018constraining}. In practice, however, these ideal conditions are unattainable, and calibration is required. 
Calibration ensures that the NRE outputs are statistically consistent and corrects for potential issues such as biases, overconfidence, or underconfidence. We follow the calibration procedure described in \citet{Brehmer_Sidd_2019_NRE, brehmer2019mining}.

NREs have been previously applied in strong lensing to infer various parameters, including substructure properties, lens profiles, and time-delay distances, directly from imaging or time-delay data \citep[e.g.,][]{Brehmer_Sidd_2019_NRE, brehmer2019mining, Coogan_2022, Eve_NRE_2023, Jarugula_2024}. 
These applications highlight the ability of NREs to extract information from complex simulations where traditional likelihood-based inference is complex or infeasible. 
In this work, we specifically extend the methodology to jointly include both subhalos and line-of-sight halos in the dark matter signal.

\section{Results}  \label{sec:results}
In this section, we present results on the sensitivity of strong lensing systems to dark matter halo populations.
We systematically examine how the number of lenses, the minimum halo mass considered, and the inclusion of line-of-sight (LOS) structure affect the constraints on DM parameters. 
Our primary goal is to assess the feasibility of excluding alternative DM models relative to the $\Lambda$CDM baseline using LSST-quality data. Throughout this section, the fiducial $\Lambda$CDM prediction is consistently indicated by the blue star in all figures. 
Note that all posterior distributions have been smoothed prior to visualization to suppress artifacts introduced by grid discretization. As explained in Appendix~\ref{app:NRE_training}, the discretization is necessary to evaluate and calibrate the trained NRE over the range of possible parameters $\vartheta$. The smoothing is done by applying a Gaussian filter over the inferred grid-based posterior, making the contours smoother, and suppressing artifacts from the discretization.

\begin{figure*}[t]
    \centering
    \includegraphics[width=0.49 \linewidth]{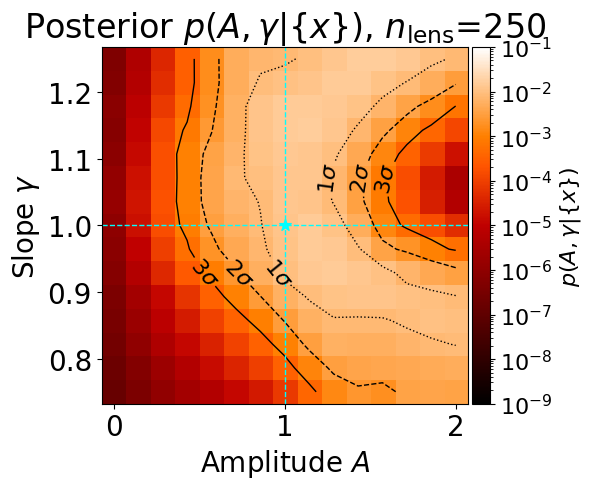}
    \includegraphics[width=0.49\linewidth]{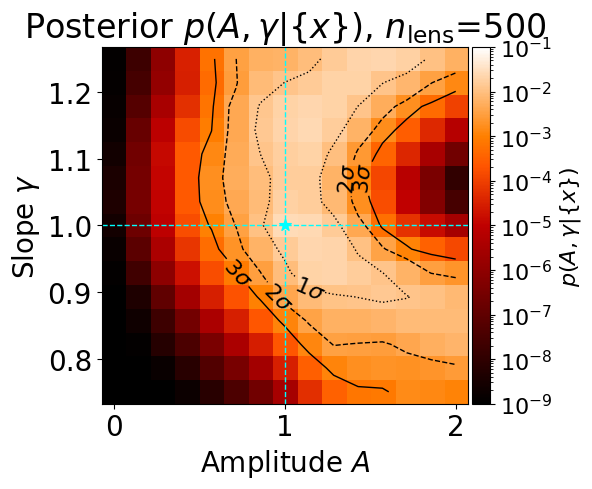}
    \includegraphics[width=0.49\linewidth]{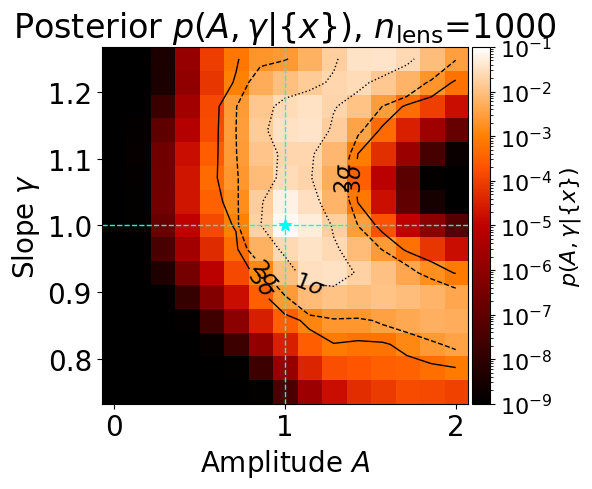}
    \includegraphics[width=0.49\linewidth]{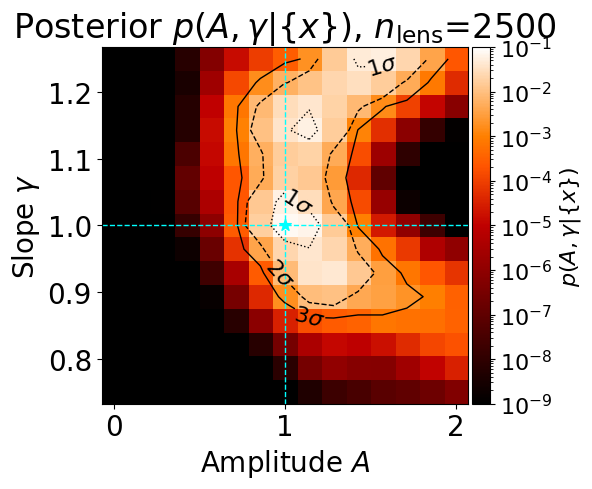}
    \caption{Posterior constraints on DM halo population parameters for 250, 500, 1\,000, and 2\,500 strong lens systems, sampled from the expected LSST redshift distribution. The contours denote $1\sigma$, $2\sigma$, and $3\sigma$ confidence regions inferred by the NRE. The blue star marks the CDM prediction.}
    \label{fig:NRE_overall_sens}
\end{figure*}

\subsection{Sensitivity Prediction as a Function of Sample Size} 
Figure~\ref{fig:NRE_overall_sens} illustrates the improvement in posterior constraints as the number of lensed systems increases from 250 to 2\,500. In each case, the lenses and sources are drawn from the approximate redshift distribution of \citet{Collett_2015}, which is binned into 70 redshift combinations in the ranges $z_{\rm l}=0.2-1.0$ and $z_{\rm s}=0.25-4.0$. 
To compute the posterior contribution of a single system, we averaged the effect of 1\,000 simulated lenses for each of the 70 redshift combinations.

The inferred posteriors reveal a strong degeneracy between high-amplitude/steep-slope mass functions and low-amplitude/shallow-slope mass functions. This could be due to similar effects on the lens of many low-mass halos vs a few higher mass halos. We leave further explorations of this degeneracy to future work.
While analyses using a few hundred lenses yield broad posteriors on the slope of the dark matter HMF comparable with other probes such as the Ly-$\alpha$ forest \citep[e.g.,][]{Viel_Lya_2005, Viel_Lya_2013, Villasenor_Lya_2023, Rogers_Lya_2025}, combining 2\,500 lenses produces significantly tighter constraints capable of excluding non-$\Lambda$CDM models. Specifically, with only 250 to 500 lenses, the constraints remain broad and do not necessarily rule out alternative DM models. However, with 2\,500 lenses - a sample size well within LSST’s expected yield - the fraction of prior volume excluded is $\approx 74\%$ at $3 \sigma$ and $\approx 36 \%$ at $5 \sigma$, narrowing the contours substantially, and enabling statistically meaningful exclusions of non-$\Lambda$CDM scenarios.

Furthermore, the ground truth is successfully contained within the $1\sigma$ posterior region. This highlights the power of large samples: the transition from hundreds to thousands of lenses marks the regime where LSST data alone can yield robust exclusion limits on dark matter models. Results obtained from lenses discovered with LSST can be used for the selection of lenses for a higher resolution follow-up to put even tighter constraints on the HMF.

\subsection{Sensitivity to Low-Mass Halos}
\begin{figure}[t]
    \centering
    \includegraphics[width=0.99\linewidth]{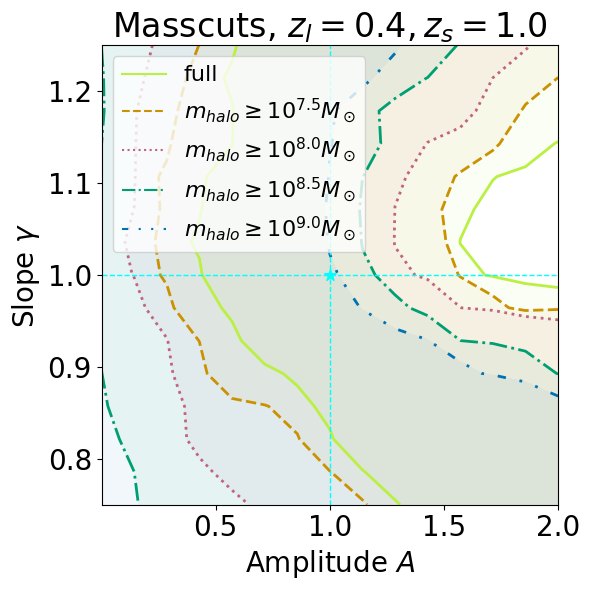}
    \caption{Posterior constraints with different minimum halo mass cuts. The contours denote the $3\sigma$ constraints from 500 lens systems. Even removing the lowest mass halos below $m_{\rm halo} \leq 7.5 \,\log_{10} M_\odot$ induces a noticeable shift relative to the full sample. Additionally, the posterior estimates seem to broaden when low-mass halos are removed.}
    \label{fig:mass_sensitivity}
\end{figure}

We next investigate whether LSST’s sensitivity arises predominantly from the high-mass end of the halo mass function or whether low-mass halos also contribute significantly. Figure~\ref{fig:mass_sensitivity} shows the posterior constraints when halos below various minimum masses are excluded for 500 lens systems. Specifically, we masked all halos with $\log_{10}(m_{\rm halo}/M_\odot) \leq 7.5, 8.0, 8.5, \text{ and } 9.0 \, M_\odot$, and compared these to the full sample which extends to a minimum mass of $\log (m_{\rm halo, min}/M_\odot)=7.0$. 
All the lens images used here for testing are the same set of lenses, with the only difference being the masking of low-mass halos during ray-tracing, which ensures direct comparability. We tested the effect of removing low-mass halos for 500 lenses, because compared to Figure~\ref{fig:NRE_overall_sens}, this test focuses on only one redshift combination.

Even modest truncation at $\log (m_{\rm halo}/M_\odot) \leq 7.5$ produces a measurable shift in the posterior, clearly demonstrating that low-mass halos contribute appreciably to the sensitivity, even for relatively low-resolution observations expected from LSST. Additionally, the posterior estimates seem to broaden a little as low-mass halos are removed. As the mass cutoff increases, the posterior broadens further, confirming that the inclusion of these small-mass halos is critical for the constraints on different dark matter models. This implies that LSST’s ability to probe or rule out the existence of these low-mass halos will directly translate into sensitivity to the fundamental properties of DM, such as the thermal velocities in WDM scenarios. In particular, the absence of halos below a given mass scale could be interpreted as evidence for suppressed structure formation, thus constraining the DM particle temperature in WDM models.

These tests were performed for a representative lens-source configuration ($z_{\rm l}=0.4$, $z_{\rm s}=1.0$), but the conclusions generalize across the full LSST redshift distribution.

\subsection{Relative Contribution of Line-of-sight and Subhalos}
\begin{figure}[t]
    \centering
    \includegraphics[width=0.99\linewidth]{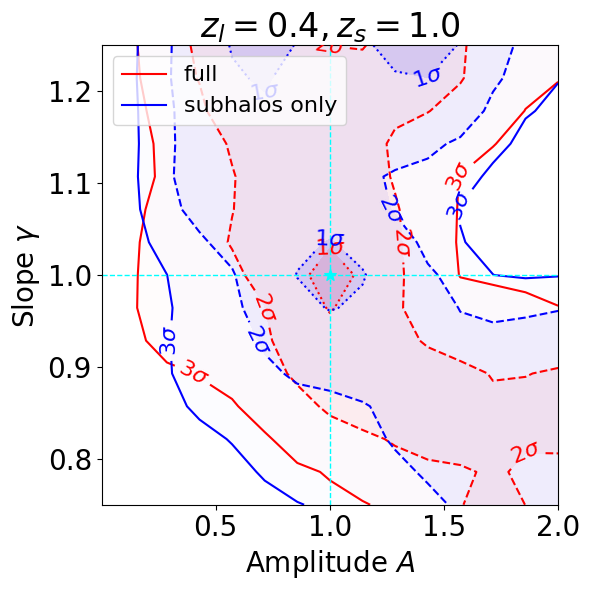}
    \caption{Posterior constraints for subhalos only (blue) compared to the full population including LOS halos (red). The contours correspond to the average $1\sigma$, $2\sigma$, and $3\sigma$ levels of combining 500 lenses. Inclusion of LOS halos significantly tightens constraints at higher and lower confidence levels.}
    \label{fig:los_vs_sub}
\end{figure}

Finally, we assess whether the overall constraints are driven primarily by subhalos in the main deflector, LOS halos, or a combination of both. To test this, we generated two matched sets of lenses: one populated with the full halo population (subhalos and LOS halos), and another where LOS halos were masked out. By using identical lenses in both cases, any difference in the resulting posteriors directly reflects the impact of LOS halos. This test was conducted on a representative lens-source configuration, $z_{\rm l}=0.4$, $z_{\rm s}=1.0$.
Specifically, we trained a new NRE independently for this redshift combination, seeing only lens systems with subhalos, as comparing with other combinations would require training a subhalo-only network for each redshift combination.

Figure~\ref{fig:los_vs_sub} compares the average posteriors derived from 500 lens samples at this representative redshift configuration. The subhalo-only case yields significantly broader contours, with the constraints from $1\sigma$ up to $3\sigma$ being consistently tighter for the full sample.
We tested the effect of subhalos against the inclusion of LOS halos in 500 lenses, because compared to Figure~\ref{fig:NRE_overall_sens}, this test focuses only on one redshift combination.
This test demonstrates that LOS halos contribute non-negligibly to the constraining power. Their importance is expected to increase with redshift, as the comoving volume and corresponding number density of line-of-sight structures grow.

These results highlight that robust population-level inference requires jointly incorporating both subhalo and LOS halo contributions. Neglecting the LOS halo population can lead to artificially weakened constraints and a potential misinterpretation of the dark matter signal.

\section{Summary and Discussion}    \label{sec:discussion}
In this work, we have quantified the sensitivity of strong gravitational lensing to dark matter halo populations using simulated lens samples representative of expected LSST observations. We explored how the overall sensitivity scales with sample size, investigated the important role of low-mass halos, and assessed the relative contributions from subhalos and line-of-sight halos. All forecasting tests were performed with a fiducial ground truth consistent with $\Lambda$CDM predictions.

Our results show that the constraining power on DM halo population parameters improves significantly with increasing numbers of strong lens systems. Posteriors inferred from samples of 2\,500 lenses yield constraints tight enough to enable the exclusion of non-$\Lambda$CDM models, highlighting the potential of LSST-quality data for probing the particle nature of dark matter. 
This trend underscores the statistical strength of large lens samples: the intrinsic degeneracy between amplitude and slope parameters in the halo mass function is progressively mitigated as more systems are added.

We also investigated whether sensitivity is driven mainly by high-mass halos or if low-mass halos measurably affect the dark matter inference. Masking halos below $\log (m_{\rm halo}/M_\odot) \leq 7.5$ produced a clear shift in the posterior, confirming that halos across the full mass spectrum contribute meaningfully to the constraints. Consequently, analyses that neglect the low-mass end risk mischaracterizing the halo population. 
In particular, non-detections of low-mass halos could provide powerful evidence for suppressed small-scale structure, placing limits on properties such as the dark matter free-streaming scale, or effective temperature, and therefore particle mass.

We further examined the respective roles of subhalos and line-of-sight (LOS) halos in shaping the posterior. By masking LOS halos in otherwise identical simulations, we found that the inclusion of LOS structure substantially tightened the constraints for all inferred posterior ranges.
This demonstrates that LOS halos are a critical component of the lensing signal, particularly at higher redshifts where their abundance increases. Consequently, future lensing analyses must account for both populations to avoid underestimating uncertainties or introducing bias into inferences.

Taken together, these findings demonstrate that LSST will enable statistically significant constraints on DM halo populations, including the power to rule out non-$\Lambda$CDM scenarios. Sensitivity arises not only from subhalos in the main deflector but also from LOS halos, and extends across the full halo mass spectrum down to our lower limit of $\sim 10^7 M_\odot$. 
This emphasizes the importance of modeling the full lensing environment, including low-mass structure and line-of-sight perturbations, when using strong gravitational lensing to probe the fundamental nature of dark matter.

Our work exclusively quantifies the possibilities to constrain the nature of DM through strong gravitational lenses with LSST data alone. The covered area, and therefore the number of lenses discovered with LSST, is unmatched and opens up unmatched opportunities despite the ground-based data quality at comparably low resolution.
Other works that study the sensitivity of the James Webb Telescope, Hubble Space Telescope, or Euclid to individual subhalo detections, subhalo mass profile slopes, and subhalo populations \citep[e.g.,][]{Riordan_2023_Euclid, Riordan_2024_HST, Riordan_2025_HST, Kollmann_2026_slope} show higher sensitivity to low-mass halos than LSST, but the strength of LSST arises through the number of systems. Combining constraints and analyses jointly from multiple surveys will lead to an even better understanding of the nature of DM.

Several limitations must be acknowledged. In this work, we assumed knowledge of the exact data-generating process and the distributions of nuisance parameters, which is an idealization not available for real observations. 
Such assumptions may lead to biases when applied to real LSST data, especially for dark matter substructure detection, where out-of-distribution effects are known to be problematic \citep[][]{OOD_paper_24}. Our results should therefore be interpreted as idealized forecasts of LSST’s potential sensitivity rather than as direct predictions for data analysis. Future work should aim to relax these assumptions by incorporating uncertainties in lens and dark matter halo modeling, imperfect lens light subtraction, source structure, and survey systematics, as well as by exploring hybrid analyses that combine LSST with complementary facilities such as \textit{Euclid} or \textit{Roman}.

In summary, strong lensing with LSST offers a powerful opportunity to probe the properties of dark matter. The large number of lenses expected will enable robust population-level inference of the halo mass function, where sensitivity is shown to depend on both subhalos and LOS halos across the entire mass range. Achieving this potential will require careful modeling of astrophysical and observational uncertainties, but the prospects for constraining or even excluding classes of dark matter models are strong.

\section*{Acknowledgments}
This paper has undergone internal review in the LSST Dark Energy Science Collaboration. 
The internal reviewers were Daniel Gilman and Supranta Boruah. 
This work is partially supported by Schmidt Science, a philanthropic initiative founded by Eric and Wendy Schmidt as part of the Virtual Institute for Astrophysics (VIA). The work is in part supported by computational resources provided by Calcul Quebec and the Digital Research Alliance of Canada. A.F. acknowledges the support from the Bourse J. Armand Bombardier. Y.H. and L.P. acknowledge support from the Canada Research Chairs Program, the National Sciences and Engineering Council of Canada through grants RGPIN-2020-05073 and 05102.

\newpage
\appendix

\section{Neural Ratio Estimator Training}\label{app:NRE_training}
\subsection{Loss Function}
Neural Ratio Estimators (NREs) are classifiers trained to distinguish between samples drawn from the joint distribution $p(x, \vartheta)$ and samples drawn from the product of marginals $p(x)p(\vartheta)$, with $x$ the samples and $\theta$ the parameters of interest.

The classifier is optimized using a modified, improved cross-entropy loss that includes a gradient regularization term, adapted from \citet{Stoye_2018} and \citet{Brehmer_Sidd_2019_NRE}. In the following, we use $z$ to denote latent nuisance parameters from the simulation, which are implicitly marginalized over to get the likelihood $p(x\mid \vartheta)$ but are necessary for computing the joint score function $t(x,z\mid \vartheta)$, i.e., the gradient of the logarithmic likelihood with respect to the parameters of interest $\vartheta$. The loss $L$ is given by
\begin{equation}
\begin{split}
    L[g(x,\vartheta)] &= \int d\vartheta \int d\vartheta' \int dx \int dz \quad \pi(\vartheta) \pi(\vartheta') p(x,z|\vartheta) \\ 
    &\times [-s \log g - (1-s) \log (1-g) - s' \log g' - (1-s') \log (1-g') \\
    &+ \alpha \left\{ \left|t\nabla_\vartheta \log \frac{1-g}{g}|_\vartheta  \right|^2 + \left|t'\nabla_\vartheta \log \frac{1-g}{g}|_{\vartheta'}  \right|^2 \right\} ]
\end{split}
\end{equation}
with
\begin{equation}
    \begin{split}
        s &= \frac{1}{1+r(x,z|\vartheta)} \\
        s' &= \frac{1}{1+r(x,z|\vartheta')} \\
        g &= g(x, \vartheta) \\
        g' &= g(x, \vartheta') \\
        t &= t(x,z|\vartheta) = \nabla_\vartheta \log p(x,z|\vartheta) \\
        t' &= t(x,z|\vartheta') = \nabla_{\vartheta'} \log p(x,z|\vartheta').
    \end{split}
\end{equation}
The joint score function $t(x,z|\vartheta)$ and the joint likelihood ratio $r(x,z|\vartheta)$ can be calculated directly when simulating each training image, with $z$ denoting the nuisance parameters in this notation.

Under perfect convergence, the optimal classifier $g_{\rm min}(x,\vartheta)$ obtained by minimizing this loss function is:
\begin{equation}
    g_{\rm min}(x,\vartheta) \equiv \arg \min L[g(x,\vartheta)] = \frac{1}{1 + r(x|\vartheta)}.
\end{equation}

The NRE prediction is $g_{\rm min}(x,\vartheta)$.
By evaluating the trained NRE $g_{\rm min}(x,\vartheta)$ at test time, we gain direct access to the otherwise intractable likelihood ratio:
\begin{equation}
    r(x|\vartheta) = \frac{1-g_{\rm min}(x,\vartheta)}{g_{\rm min}(x,\vartheta)} = \frac{p(x\mid \vartheta)}{p(x)}.
\end{equation}
This ratio is useful when we consider Bayes’ rule:
\begin{equation} \label{eq:bayes_rule}
    p(\vartheta \mid x) \;=\; \frac{p(x \mid \vartheta) \, p(\vartheta)}{p(x)}.
\end{equation}
With the learned ratio and the prior $p(\vartheta)$ on the parameters, one can recover the approximate posterior:
\begin{equation}
    p(\vartheta | x) \;\propto\; r_{\phi}(x,\vartheta)\,p(\vartheta).
\end{equation}

We use the Adam optimizer \citep{Adam_optim_2014} with an initial learning rate of $10^{-4}$ and a scheduler set to decay linearly over 75 epochs to a final learning rate of $10^{-6}$.
The gradient regularization coefficient is set to $\alpha = 10^{-5}$.
Our NRE architecture is based on a ResNet18 \citep{ResNet_2016}, coupled with two fully connected layers. The hidden layer has a dimensionality of 512.

\subsection{Calibration}
We employ a histogram-based calibration procedure, following \citet{Cranmer_2015} and \citet{brehmer2018constraining}.

For every parameter point of interest $\vartheta$, we simulate a set of images $\mathcal{X}_{\vartheta} = \{x_j\}$ drawn from the conditional distribution $p(x|\vartheta)$. For each image in this set, we calculate the raw network prediction $\hat{r} = r(x_j \mid \vartheta)$.
Additionally, we generate a reference set of images drawn from the marginal distribution $p_{\rm ref}(x) = p(x)$ (by marginalizing over the prior and drawing randomly from the full prior range) and calculate their corresponding network predictions.

The calibrated likelihood ratio for a specific observation $x$ and parameter $\vartheta$ is given by the ratio of the empirical densities of the network outputs:
\begin{equation}
    r_{\cal}(x \mid \vartheta) = \frac{\hat{p}(\hat{r}\mid \vartheta)}{\hat{p}_{\rm ref}(\hat{r})}
\end{equation}
where $\hat{r} = r(x|\vartheta)$ is the raw network output, and $\hat{p}(\cdot)$ is the probability density approximated through the histograms.

For calibration, we generate for each single NRE 15\,000 strong lenses for the reference distribution, and 5\,000 systems for each possible considered bin of HMF parameter combination.

While the computational costs of this calibration scale with the number of evaluated parameter points, the calibration ensures the reliability of the inference.
The inference results are conservative and will not be overconfident, even if the uncalibrated output deviates from the true likelihood ratio \citep[][]{Brehmer_Sidd_2019_NRE}.

\bibliography{sample631}{}
\bibliographystyle{aasjournal}

\end{document}